\documentclass[aip,pop,
 amsmath,amssymb,
 reprint,%
]{revtex4-2}
\usepackage{float}
\usepackage{graphicx}% Include figure files
\usepackage{dcolumn}% Align table columns on decimal point
\usepackage{bm}% bold math
\usepackage{xcolor}
\usepackage[utf8]{inputenc}
\usepackage[T1]{fontenc}
\usepackage{mathptmx}
\usepackage{ulem}
\usepackage{etoolbox}
\usepackage{hyperref}
\usepackage{xcolor}
\makeatletter
\def\@email#1#2{%
 \endgroup
 \patchcmd{\titleblock@produce}
  {\frontmatter@RRAPformat}
  {\frontmatter@RRAPformat{\produce@RRAP{*#1\href{mailto:#2}{#2}}}\frontmatter@RRAPformat}
  {}{}
}%
\makeatother
\begin{document}

\preprint{AIP/123-QED}

\title{Investigating Solid-Fluid Phase Coexistence in DC Plasma Bilayer Crystals: The Role of Particle Pairing and Mode Coupling}
% Force line breaks with \\
\author{Siddhartha Mangamuri}
\affiliation{Department of Physics, Indian Institute of Science Education and Research, Dr. Homi Bhabha Road, Pune 411008, India}%Lines break automatically or can be forced with \\

\author{L Cou{\"e}del}

\affiliation{Department of Physics and Engineering Physics, University of Saskatchewan, Saskatoon, Saskatchewan S7N 5E2, Canada}%
\affiliation{CNRS, Aix-Marseille Université, Laboratoire PIIM, 13397 Marseille, France}

\author{S. Jaiswal}%
\email{surabhi@iiserpune.ac.in}
%\homepage{http://www.Second.institution.edu/~Charlie.Author.}
\affiliation{Department of Physics, Indian Institute of Science Education and Research, Dr. Homi Bhabha Road, Pune 411008, India}%\\This line break forced with \textbackslash\textbackslash
\date{\today}% It is always \today, today,

\begin{abstract}

This article presents a detailed investigation of solid-fluid phase coexistence in a bilayer dusty plasma crystal subjected to varying confinement 
ring bias voltages in a DC glow discharge argon plasma. Melamine formaldehyde particles were employed to form a stable, hexagonally ordered bilayer 
crystal within a confinement ring electrically isolated from the grounded cathode. By systematically adjusting the confinement ring bias, a distinct 
phase coexistence emerged characterized by a fluid-like melted core surrounded by a solid crystalline periphery. Crucially, analysis of the phonon spectra 
revealed frequency shifts that deviate significantly from the predictions of classical monolayer Mode-Coupling Instability (MCI) theory. 
Stability analysis further demonstrated that dynamic interlayer particle pairing and the associated increase in non-reciprocal interaction strength are strongly correlated with the onset of structural destabilization. These findings highlight previously underappreciated mechanisms driving the melting transition in bilayer dusty plasmas, 
offering a more comprehensive understanding of phase behavior in complex plasma systems. The results underscore the importance of interlayer 
coupling and confinement effects in tuning structural transitions.

\end{abstract}

\maketitle
\section{\label{sec:level1}Introduction}
Dusty plasma crystals, consisting of monodisperse micrometer-sized particles suspended in a gas discharge plasma, provide a unique platform to study phase transitions~\cite{PhysRevE.53.2757,Jaiswal2019,PhysRevLett.80.5345,Naumkin2018,Jaiswal2015,Jaiswal2017}, stability~\cite{Couedel2010,PhysRevLett.119.255001}, 
and melting dynamics in systems with long-range interactions \cite{Thomas1996}. Micron or sub-micron sized particles once inserted into the plasma, 
become highly charged due to interactions with the surrounding ions and electrons and form stable, lattice-like 
structures under suitable confinement~\cite{Chu1994,Thomas1994,FORTOV20051}, giving rise to what is termed a “plasma crystal.” 
Such systems are analogous to soft condensed matter systems and offer a controlled means of investigating particle-particle 
interactions, solid-fluid coexistence, and melting processes, making them valuable for studies that bridge plasma physics 
and condensed matter theory~\cite{HAMAGUCHI199957,RevModPhys.81.1353}. 
\par In recent years, dusty plasma crystals have emerged as a model system for exploring phase coexistence~\cite{Singh2023,Jaiswal2024,Binder03072021,Dickman2016,LUKACS20222230,PhysRevLett.110.055701,C7SM01504F}, where solid and fluid phases can coexist within the same structure. The dynamics underlying this phenomenon are complex, as they involve particle interactions, structural rearrangements, and instabilities that are not commonly observed in simpler systems. Phonon modes play a critical role in understanding phase transitions and melting dynamics in complex plasmas.  
Longitudinal and transverse phonon modes, arising from compressional and shear motions respectively, provide  
insights into system stability and energy distribution~\cite{Wang2001,Nunomura2002,Liu2003}. Wang  
\textit{et al.}~\cite{Wang2001} developed a theoretical framework for phonon modes in Yukawa crystals, a model  
analogous to dusty plasmas with screened Coulomb interactions. This enabled systematic studies of dispersion  
relations and the influence of confinement and interaction strength on phonon behavior and instabilities~  
\cite{Ivlev2002,Zhdanov2009,Couedel2010,Couedel2022}. 

\par In bilayer or multilayer complex plasma crystals, the Schweigert instability is a commonly observed  
phenomenon that results in the melting of the crystal structure~\cite{PhysRevLett.80.5345,Schweigert1998a,  
Schweigert2000}. It originates from the non-reciprocal nature of interactions between particles in adjacent  
layers due to constant ion flows, which produce positively charged regions ion wakefields downstream of each  
particle~\cite{Melzer2000a,Kompaneets2007,Kompaneets2016}. These ion wakes exert attractive forces on nearby  
particles, rendering the microparticle interactions non-reciprocal. In multilayer systems, ion wakes of  
upper-layer particles attract lower-layer particles, but no corresponding force acts in the opposite direction.  
This asymmetry leads to a non-reciprocal force balance that destabilizes vertical particle alignment and  
results in growing horizontal oscillations a process identified as the Schweigert instability that can  
ultimately melt the ordered plasma crystal~\cite{Singh2023}. 
Zhdanov \textit{et al.}~\cite{Zhdanov2014,Zhdanov2015}  
reported spontaneous particle pairing and collective dynamics in two-dimensional plasma crystals, showing that  
the accumulation of such pairs contributes to structural melting.

\par Studies on phase transitions in monolayer complex plasmas have shown that mode-coupling instability (MCI) is the primary mechanism responsible for melting. ~\cite{Couedel2009,Couedel2010,  Zhdanov2009,Zhdanov2015,Couedel2022}. It was observed that the coupling between longitudinal and out-of-plane modes triggers the destabilization of the lattice, ultimately leading to melting. Ivlev \textit{et al.}~\cite{Ivlev2017} later proposed a  
common framework for monolayer and bilayer crystals, highlighting wake-mediated non-reciprocity, MCI and 
a universal stability principle for both monolayer and bilayer complex plasmas. The instability is driven by the 
coupling of wave modes, which is caused by non-reciprocal, wake-mediated interparticle forces. In monolayers, 
ion wakes introduce a weak negative coupling between the normally separate horizontal and vertical modes. Conversely, 
bilayers possess an inherent positive coupling between horizontal modes due to their structural asymmetry. Therefore, 
for a bilayer to become unstable, the wake-induced negative coupling must be strong enough to overcome this existing 
positive offset, a condition that is facilitated in both systems when modes are in resonance. Recently, Jaiswal  
\textit{et al.}~\cite{Jaiswal2024} reported the first experimental observation of phase coexistence in bilayer  
dust clouds, induced by confinement variations. While they observed signs of MCI, the absence of   
intersection between vertical and horizontal modes and phase coexistence seemed to emerge when  
interlayer and interparticle spacing approach or fall below the screening length. This highlights the need  
for detailed investigation of wake-mediated non-reciprocal interactions and their role in bilayer crystal  
stability.

This article investigates the mechanisms underlying phase coexistence and melting in bilayer dusty plasma crystals, 
with a specific focus on the interplay between particle pairing and mode coupling under varying confinement ring bias 
voltages in a DC glow discharge plasma. By analyzing phonon spectra across a range of confinement potentials, we explore 
how longitudinal and transverse modes respond to particle interactions and coupling effects. Through detailed tracking of particle trajectories, separations, and velocities, we uncover new insights into the dynamic processes that destabilize the crystal structure. Our results demonstrate that dynamic particle pairing, combined with a modified form of MCI, is strongly correlated with the observed structural transition of bilayer dust crystal. Additionally, we introduce and apply a quantitative analysis of interlayer non-reciprocity, 
capturing the emergence of asymmetric particle interactions that signal structural instability. This approach provides a 
novel diagnostic for identifying the onset of disorder and melting in bilayer dusty plasmas. Overall, this study contributes to a deeper understanding of structural stability and phase transitions in complex plasma systems, offering new perspectives 
on the microscopic interactions that govern these phenomena.

\begin{figure}[htbp]
   \centering
    \includegraphics[width=\linewidth]{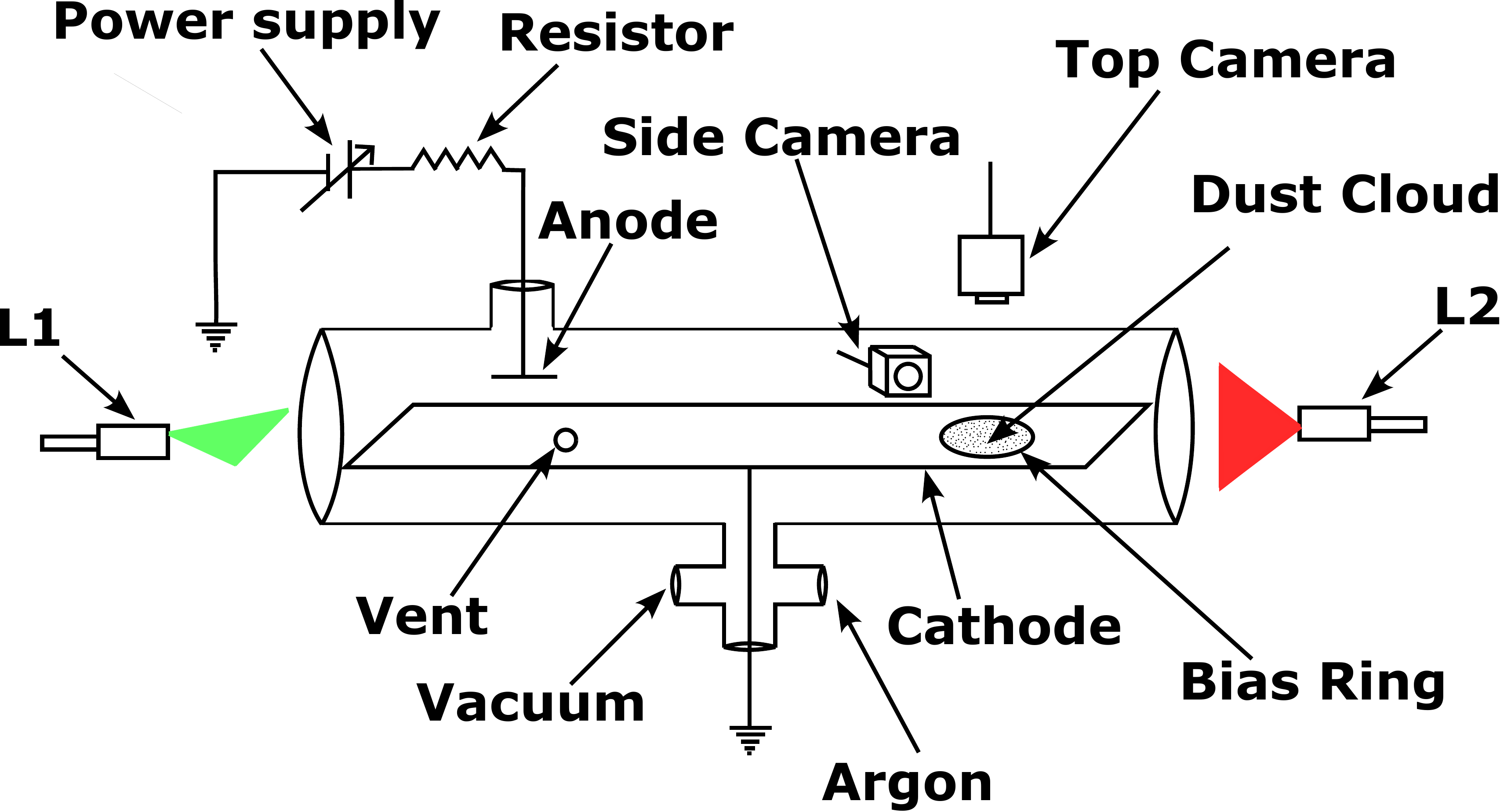}
    \caption{Schematic of the experimental arrangement. L$_1$ and L$_2$ are green (532 nm) 
    and red (650 nm) illumination lasers. }
   \label{experimental setup}
\end{figure}

\begin{figure}[htbp]
        \centering
        \includegraphics[width=0.9\linewidth]{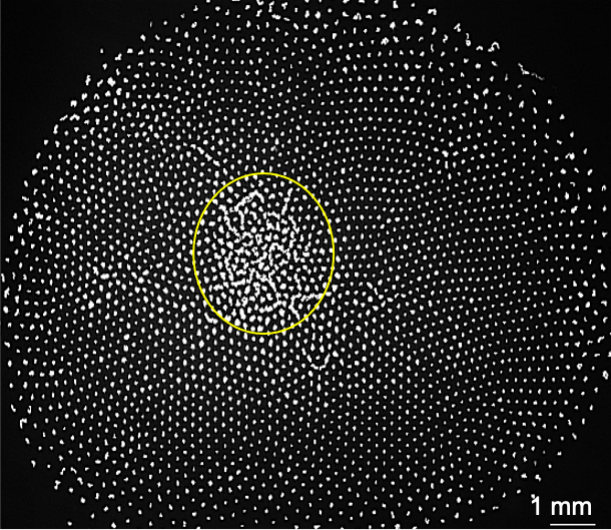}
        \caption{Snapshot showing overlapped
image of 20 consecutive frames from the top view camera at 103 V of confining ring voltage. Circular marked region shows the region of melting. The outside region is almost stationary with a small thermal fluctuation around the equilibrium position.}
        
        \label{topview}
\end{figure}

\begin{figure}[htbp]
    \centering
        \includegraphics[width=1\linewidth]{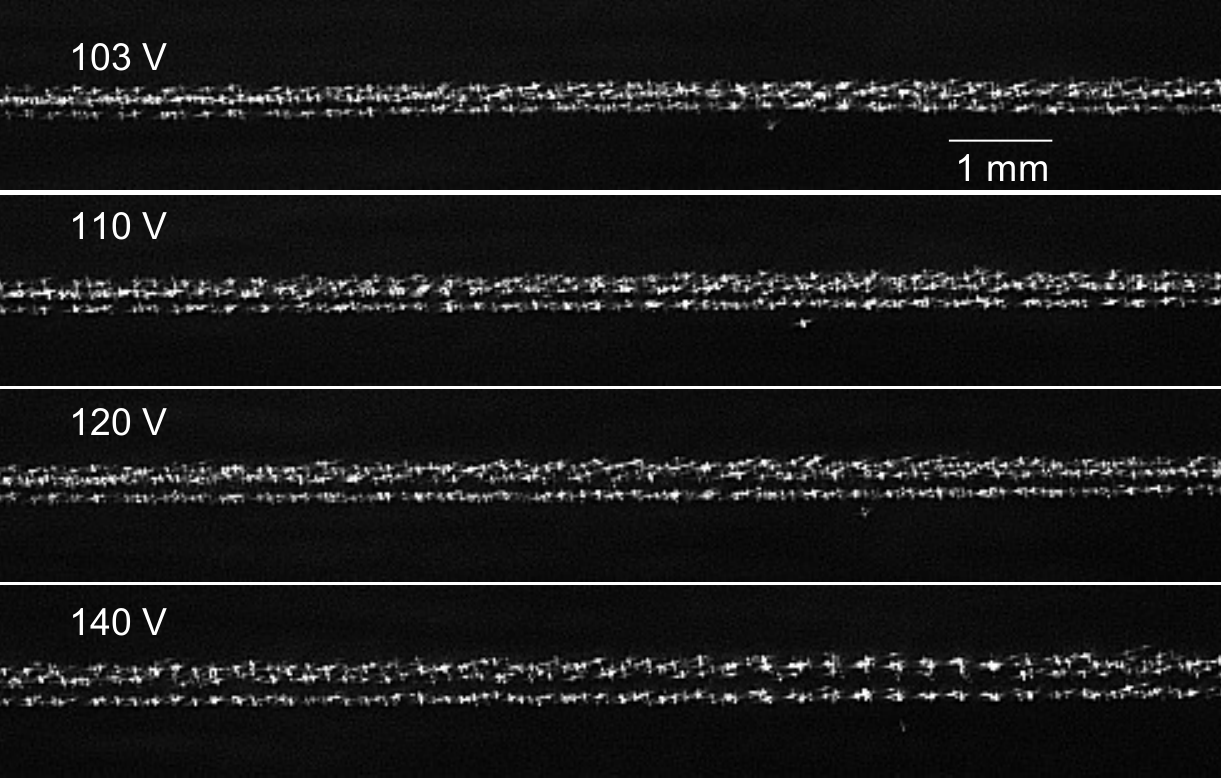}
        \caption{View of the dust cloud from the side camera at different bias voltage. The layer gap decreases with reducing confinement potential from 140 V to 103 V. }
        \label{sideview}
\end{figure}
%%%%%%%%%%%%%%%%%%%%%%%%%%%%
\section{\label{sec:level2}Experimental procedure}

The experimental setup closely follows the configuration described by Jaiswal \textit{et al.}~\cite{Jaiswal2024} in the 
Low-Temperature Dusty Plasma Experimental device (LDPEx). A schematic of the setup is shown in Fig.~\ref{experimental setup}. 
The gas discharge was initiated in a borosilicate glass tube measuring 33~cm in length and 7.6~cm in inner diameter. 
The tube is equipped with multiple axial and radial ports to enhance accessibility and diagnostic capabilities. 
A stainless-steel (SS) disk anode (2.5~cm diameter) and a 30~cm~$\times$~6~cm~$\times$~0.6~cm SS plate cathode were positioned 
6~cm apart inside the chamber. This asymmetric electrode configuration minimizes ion heating of the dust particles, enabling 
the formation of a stable dust plasma crystal in the DC glow discharge environment. The chamber was initially pumped down to 
a base pressure of $10^{-3}$~mbar, then filled with argon gas to maintain a working pressure of 112~mTorr (15~Pa). 

To initiate plasma, a DC power supply applied a voltage of 350--400~V across a 2~k$\Omega$ resistor connected in series with 
the anode, resulting in a discharge current of 2--5~mA. Both the argon pressure and discharge voltage were kept constant throughout 
the experiment. The dust particles used were monodisperse melamine-formaldehyde spheres with a diameter of $7.14 \pm 0.06~\mu$m. 
Upon introduction into the plasma, the particles acquired negative charges and levitated within the plasma sheath above the cathode.
Radial confinement of the dust particles was provided by an aluminum ring isolated from the grounded cathode using a ceramic cover. 
The bias potential on the ring was adjustable and varied from 100~V to 140~V during the course of the study. This variation modified 
the sheath structure, thereby inducing structural transitions within the dust crystal.
A 532~nm, 100~mW laser (L1) was expanded horizontally into a thin sheet to illuminate a specific horizontal layer of the dust crystal. 
A second laser (L2, 650~nm, 100~mW) was expanded vertically to illuminate a vertical cross-section of the particle cloud in the X-Z plane. The top and side view camera was used to record the respective images in horizontal and vertical direction enabling the complete picture of particle dynamics in our system. The top-view camera was focused such that only the uppermost particle layer was visible.

\begin{figure*}[htpb]
    \centering
    \includegraphics[width=\textwidth]{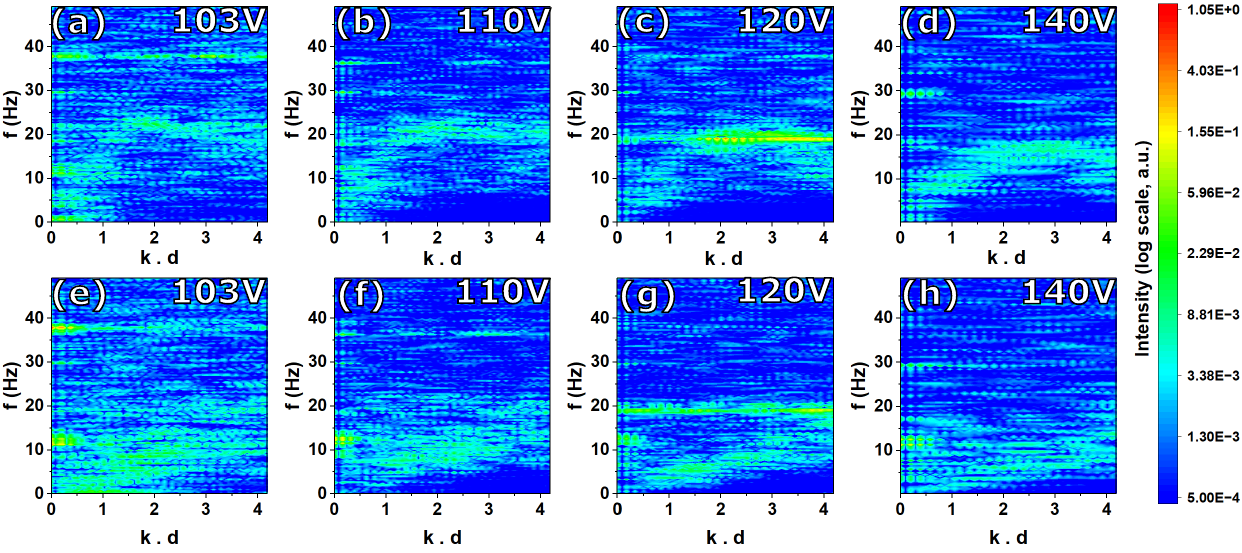}
    \caption{Longitudinal (top) and transverse (bottom) spectra for $0^\circ$ k-vector orientation 
    }
    \label{acoustic0}

    \vspace{0.5cm} % adjust spacing between the two plots

    \includegraphics[width=\textwidth]{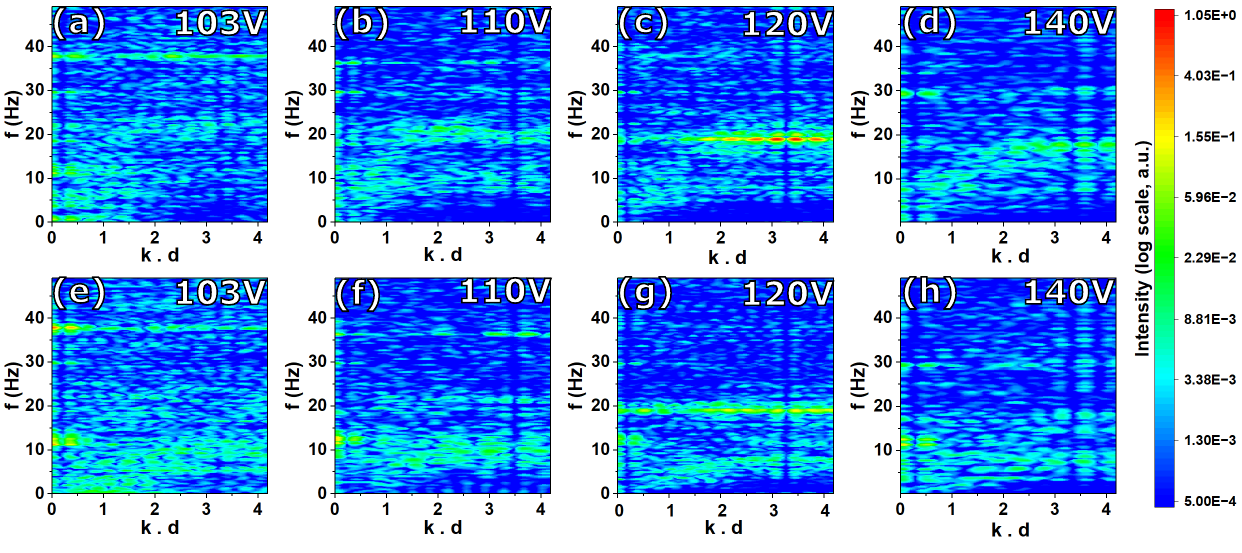}
    \caption{Longitudinal (top) and transverse (bottom) spectra for $30^\circ$ k-vector orientation 
    }
    \label{acoustic30}
\end{figure*}
%%%%%%%%%%%%%%%%%%%%%
\begin{figure*}
    \centering    
    \includegraphics[height=0.22\textwidth,width=\textwidth]{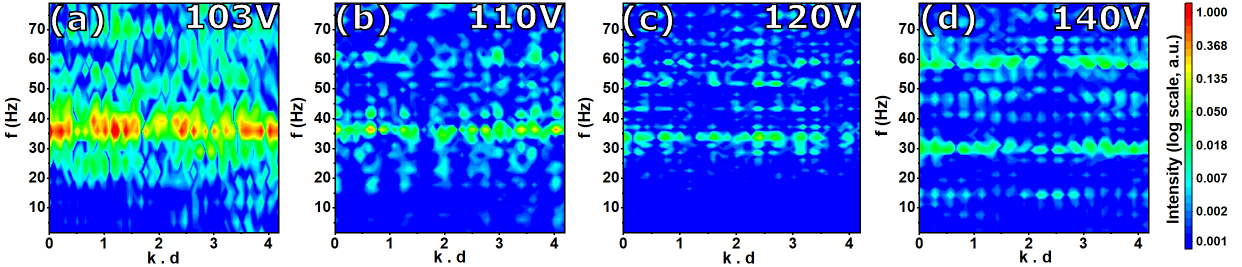}
    \caption{Optical phonon modes for different bias voltages
    }
    \label{optical modes}
\end{figure*}

%%%%%%%%%%%%%%%%%%%%%%%%%%%%%%%%%%%

\section{\label{sec:level3}Results}
Fig.~\ref{topview} shows an overlapped image of 20 consecutive
frames from the top view camera at 103 V of confining ring voltage. The circular marked region shows a clear melting whereas the periphery is almost stationary with a small thermal fluctuation around the equilibrium position. The side view of the particle cloud (shown in Fig.~\ref{sideview}) reveals that the interlayer gap between two layers is decreasing as the confinement potential decreases from 140 V to 103 V. This increases the interaction between the layers and vertical oscillations and the coupling of the two layers was seen for the voltage below 120 V (see supplementary material).

Clear vertical oscillations in both layers suggest that the Schweigert mechanism may not be directly applicable to our bilayer system. Moreover, this mechanism assumes a damping-strength threshold beyond which the crystal melts, whereas in our case the damping strength (i.e., neutral pressure) remains constant. Our results indicate that structural changes are mainly driven by variations in the confining potential, so the damping threshold appears to play little role in this context.
Side-view imaging provides further insight into the structural evolution of the bilayer with decreasing confinement bias. At high bias voltages (140 V), the system forms a stable bilayer with small-amplitude vertical oscillations in both layers. As the bias voltage is reduced toward 120 V, vertical oscillations become more pronounced in both layers (see Supplementary Figs. 2 and 3), indicating increased coupling to the sheath electric field. Upon further reduction of the bias voltage, the upper layer intermittently splits into two vertically separated sublayers, while the lower layer continues to oscillate as a single layer. The splitting of the upper layer causes the reduced layer gap and enhances both interlayer and intralayer interactions. The combined effect of strengthened vertical oscillations in both layers and the fragmentation of the upper layer promotes particle pairing and non-reciprocal energy transfer, ultimately leading to the melting observed in the top-view measurements.

\subsection{Phonon spectra}\label{section:modecoupling}

The phonon spectrum of the dust lattice~\cite{Couedel2009,Zhdanov2014} was analyzed over a range of confinement ring  
bias voltages, focusing on optical modes (from side-view data) and acoustic modes (from top-view data). Particle  
positions were extracted using ImageJ~\cite{Schneider2012} for both views. 
 
The extracted positions were processed using the Trackpy library~\cite{Couedel2019,Allan2025}  
to link particle trajectories and compute velocities. Longitudinal and transverse phonon currents were then calculated  
following the method of Nunomura \textit{et al.}~\cite{Nunomura2002}, and Fourier transforms were applied to obtain the current fluctuation  
spectra. This analysis across various ring bias voltages enabled us to study the evolution of wave modes in the crystal.

As the confinement bias decreased from 140~V to 103~V, we observed a notable increase in the maximum frequencies of both  
longitudinal and transverse acoustic modes, rising from 17.5~Hz to 22.5~Hz (Figs.~\ref{acoustic0},~\ref{acoustic30}).  
Similarly, optical mode frequencies increased from approximately 30~Hz at 140~V to around 39~Hz at 103~V  
(Fig.~\ref{optical modes}), a trend also visible in the normalized power spectral density~\cite{Jaiswal2024}. This  
upward frequency shift suggests that reducing the ring bias voltage stiffens the confinement potential, thereby  
strengthening interparticle interactions and enhancing collective dynamics.

At a confinement bias of 120~V, signs of torsional motion~\cite{Nosenko2017} become visible, indicating the onset of  
particle pairing [see hotspots in Figs.~\ref{acoustic0}(c), (g) and~\ref{acoustic30}(c), (g)]. As previously reported, torsion and MCI develop over similar spatial and temporal scales~\cite{Nosenko2017}. However, in contrast to monolayer MCI, no localized hot spots are present in the out-of-plane spectra (Fig.~\ref{optical modes}(c)). Below 120~V, spectral features appear increasingly blurred. It is possible that the imaging setup captured particles from both layers, making the linking of particle trajectories inaccurate and leading to artificial mixing of spectral features. This behavior likely reflects a change in interparticle coupling, possibly due to the formation of particle pairs between the two layers. 

Such pairing can modify the system’s vibrational response and affecting the  
mechanical stability of the lattice~\cite{Nosenko2017}.
  
The out-of-plane spectra show an enhancement in vertical motion (Fig.~\ref{optical modes}), akin to classical monolayer  
MCI. \cite{Couedel2009,Couedel2010} Furthermore, traces of the out-of-plane mode becomes visible in the in-plane spectra around 35--40~Hz for voltages below 120~V,  
suggesting interlayer mode coupling. While some features such as acoustic mode resonances resemble traditional MCI,  
the absence of a corresponding optical hot spots and the presence of harmonics point to a modified instability.  
This bilayer-specific variant likely stems from vertical asymmetry in the interaction potential and dynamic pairing  
behavior. These harmonics support the view that vertical coupling and interlayer asymmetry significantly reshape  
phonon mode structure, resulting in modified instability mechanisms unique to bilayer complex plasma systems.

\subsection{Dependence of particle pairing on the confining ring bias voltage}

\begin{figure}[htbp]
    \centering
    \includegraphics[width=0.35\textwidth]{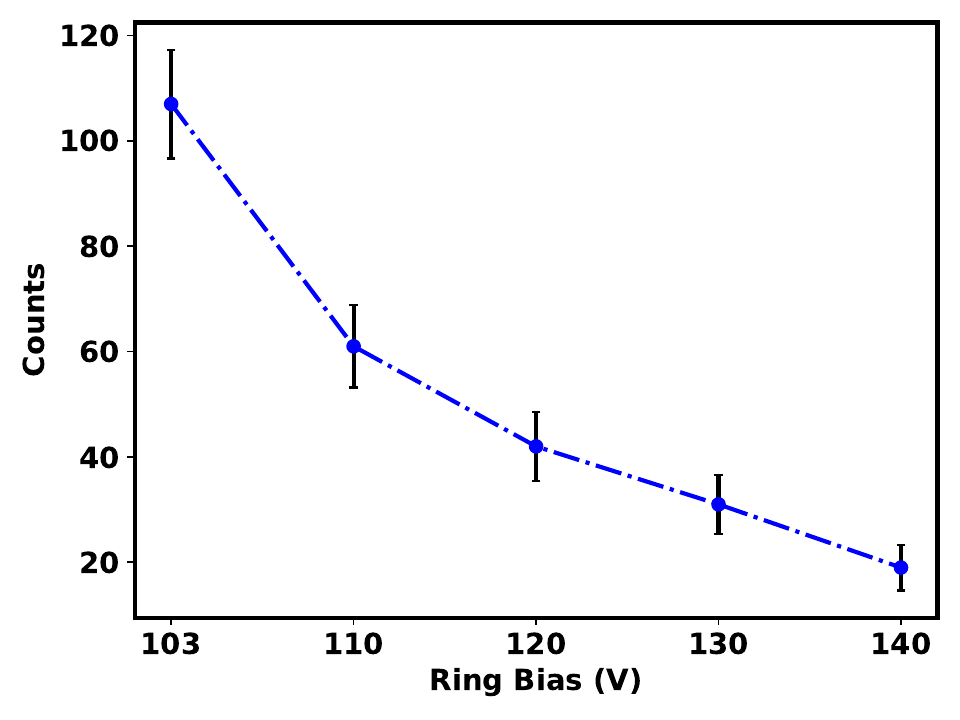}
    \caption{Evolution of the number of particle pairs observed as function of the confining ring bias voltage. Pair counts were performed manually on top-view frames; uncertainty is reported as the counting (Poisson) error $\sqrt{N}$ and conservatively covers human error in identifying overlapping/masked particles.}
    \label{numpairs_vs_voltage}
\end{figure}

As the ring bias voltage was reduced, a noticeable increase in the number of particle pairs formed was observed. Tracking the number of pairs as a function of voltage 
revealed a clear trend: the number of pairs increased as the voltage decreased (Fig.~\ref{numpairs_vs_voltage}), 
particularly around 120~V and below. At 103 V, the number of pairs in the field of view was measured as 107 ± 10, confirming the strong increase in pair formation at lower confinement potentials

\par At higher ring bias voltages, such as 140~V, particle interactions were weaker, and pair formation occurred less frequently. As the voltage decreased, the frequency of pairing events increased, indicating stronger interactions. This observation 
aligns with the phonon spectrum data, where the maximum phonon frequency increased with decreasing voltage, 
reflecting a more strongly coupled system. The intensifying interactions at 
lower voltages appear to destabilize the structure, pushing the system toward a more fluid-like state.

Overall, the relationship between voltage and pair formation underscores the critical role of particle 
pairing in the transition from a solid-like phase to a more disordered, fluid state.

\subsection{\label{casestudy}Case study: dynamics of particle pairing at a confining ring bias voltage of 120~V}

As we have seen in previous sections, at 120~V, significant changes in particle behavior were observed. For example, the hot spot arises at 120 V and the pairing effect between the particles also increases drastically after 120 V. This has important consequence for particle structural dynamics therefore, 120 V has been considered as a critical voltage for understanding the transition from crystalline to fluid states.

\begin{figure}[htbp]
    \centering
    \includegraphics[width=0.3\textwidth]{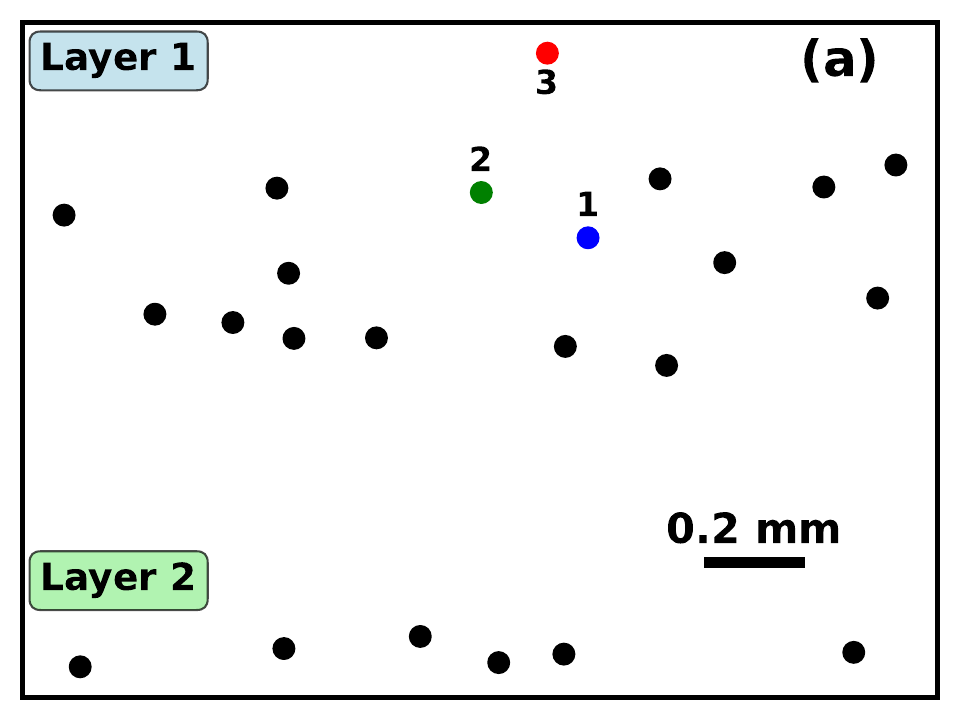}
    \includegraphics[width=1.0\linewidth]{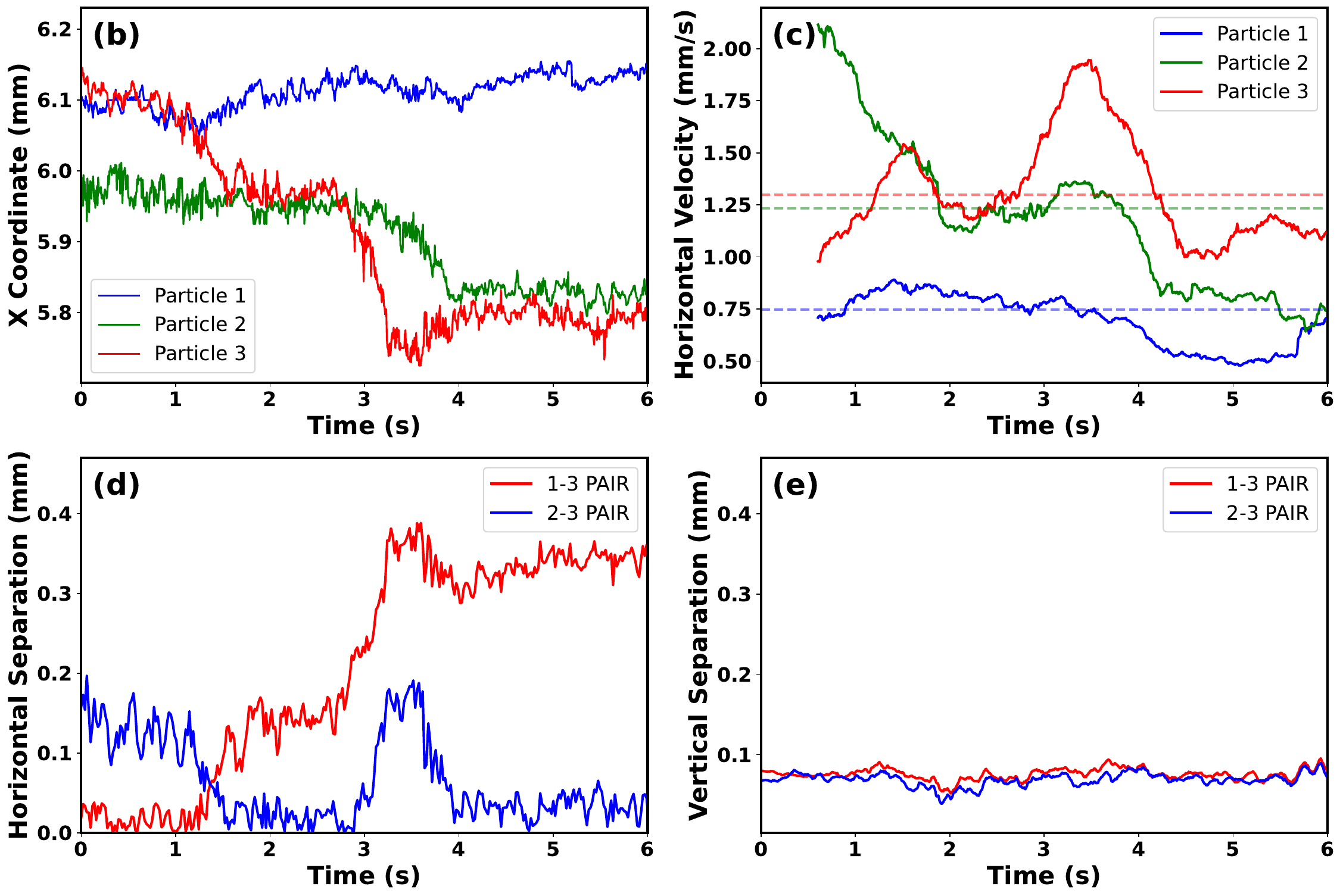}
    \caption{(a) Coordinates of particles in the region of interest. 
    (b) Horizontal trajectory of particles 1 (blue), 2 (green), and 3 (red). 
    (c) Fluctuations in horizontal velocity of particles. The dashed lines represent the average horizontal velocity of each particle. 
    (d) Horizontal separation between particles in both pairs. 
    (e) Vertical separation between particles in both pairs. }
    \label{fig_pairing_120V}
\end{figure}

In order to illustrate that, we focus on three particles: "1" and "2" in the top layer (Layer 1 in Fig.~\ref{fig_pairing_120V}(a)), 
and "3" just above Layer 1. Their motion over 6 seconds reveals the dynamics of pair formation, collapse, 
and particle dragging, contributing to system destabilization. Fig.~\ref{fig_pairing_120V}(b) shows 
the evolution of their $x$-coordinates.
Initially, particle "3" formed a stable pair with "1" for $\sim$1.5~s, as shown by their nearly identical 
$x$-coordinates. At $t \simeq 1.5$~s, the pair broke, and "3" drifted 
leftward toward particle "2," forming a new pair shortly before $t = 2$~s. Around $t \simeq 3$~s, 
particle "3" continued leftward and while dragging along particle "2" which gradually moved leftwards too. 
The pair realigned by $t \simeq 4$~s. This dragging is the signature of the ion wake below particle "3" 
influences nearby particles (also evidenced by Melzer \emph{et al.}\cite{Melzer2000a}), 
potentially destabilizing the entire system.

Fig. \ref{fig_pairing_120V}(c) represents the horizontal velocity fluctuations which provide a direct measure of energy exchange during the pairing process. The average horizontal velocities were $\langle V_{x_1}\rangle$ = 0.747 mm/s, $\langle V_{x_2}\rangle$ = 1.234 mm/s, and $\langle V_{x_3}\rangle$ = 1.298 mm/s, indicating that particle 3 exhibits the strongest horizontal motion, consistent with its role in successive pair interactions. 
Distinct velocity spikes in particle 3 occur during its transition from pairing with particle 1 to forming a pair with particle 2, as well as during the intermediate phase where it drags particle 2 signifying a moment of energy transfer and structural rearrangement. A corresponding peak in the velocity of particle 2 confirms their coupled motion. Once particle 3 establishes a stable pairing with particle 2, the velocities of both particles stabilize, indicating the re-establishment of a transient equilibrium. These correlated variations in $V_{x}$ highlight the dynamic exchange of momentum during pair formation and collapse, central to the onset of local disorder in the bilayer.

Horizontal separation as shown in Fig.~\ref{fig_pairing_120V}(d) remained constant for the first 1.2~s, then diverged:
\begin{itemize}
    \item 1-3 pair: separation increased after 1.2~s, stabilized around 3.9~s.
    \item 2-3 pair: separation decreased after 1.2~s, constant until 3~s, then increased at 3.6~s as "3" moved 
    leftward. It decreased again as "2" was pulled left, stabilizing around 4.2~s.
\end{itemize}
Vertical separation [Fig.~\ref{fig_pairing_120V}(e)] remained mostly constant at $0.06 \pm 0.01$ mm 
% \textcolor{red}{error seems oddly large here.}\textcolor{blue}{There was typo. The error is 0.01 mm}
, indicating that the  dragging was primarily 
horizontal. These trends reflect how particle "3" influenced the motion and separation of particles 
"1" and "2" over time.

The horizontal dragging force of particle "3" on particle "2" was estimated as:
\[
F_x = \gamma \cdot \langle V \rangle \cdot m = 0.413~\text{fN}
\]
where $\gamma$ is the Epstein drag coefficient~\cite{espteindrag}, $\langle V\rangle$ is the average horizontal velocity and $m$ is the mass of the particles. 

Assuming the system is overdamped and using $F_x = k \cdot \Delta x $, the spring constant for the 2-3 pair during the interaction was:
\[
k = 3.75~\text{pN}\cdot\text{m}^{-1}
\]

In this expression, $\Delta x$ is defined as the instantaneous horizontal displacement of the particle from its equilibrium position, corresponding to the effective lateral deformation produced by the wake-induced interparticle attraction. The estimated horizontal dragging force and effective spring constant quantify the strength of the wake-mediated coupling between paired particles. These values confirm that the interparticle attraction is sufficiently strong to produce measurable lateral displacement and energy exchange. The spring constant thus characterizes how mechanical energy is stored and released during pairing events, linking microscopic force asymmetry to the observed velocity fluctuations and, ultimately, to the onset of melting.~\cite{Zhdanov2015}.

The dragging of "2" by "3" provides key insight into the melting transition. Increasing interaction strength, 
reduced particle distance, and energy transfer from "3" disrupt the crystal structure. Continuous pair 
formation and collapse, such as observed between particles "2" and "3", drive the system toward a fluid-like state.
Velocity fluctuations and changing horizontal separation suggest that energy transfer during pairing events 
is a primary driver of melting. These dynamic interactions locally destabilize the structure, leading to 
crystallinity breakdown.

\subsection{non-reciprocal interactions and driving strength}

In complex plasmas, the effective interaction between microparticles is mediated by streaming ions. The resulting wakefields produce an inherently anisotropic potential that depends on flow direction and spatial asymmetry. In bilayer systems, this anisotropy translates into a fundamental breakdown of Newton’s third law: the attractive force experienced by a lower-layer particle due to the ion wake of an upper-layer particle is not balanced by a corresponding force in the opposite direction. This nonreciprocity makes bilayer crystals qualitatively different from equilibrium Yukawa systems, as the microscopic forces themselves carry a directional bias that drives the system away from detailed balance.

To quantify this effect, we introduced a pair-resolved metric following from the recent work from Ivlev et al. \cite{Ivlev2015}
\begin{align}
    \mathbf{R}=<|F_{top}+F_{bottom}|>
\end{align}

Here $F_{top}$ and $F_{bottom}$ denote the instantaneous forces acting on the upper and lower layer particles within a pair, respectively, as obtained from their measured accelerations using the particle trajectories. This directly probes the imbalance of forces in paired particles across the two layers. In the case of reciprocal interactions, forces cancel and 
\textbf{R} vanishes, whereas any finite value signals that wake-mediated asymmetry injects net momentum into the system. This measure has the advantage of being experimentally accessible from trajectory data, avoiding the need for coarse-grained approximations or indirect stability proxies.

\begin{figure}[htb!]
    \centering
    \includegraphics[width=1\linewidth]{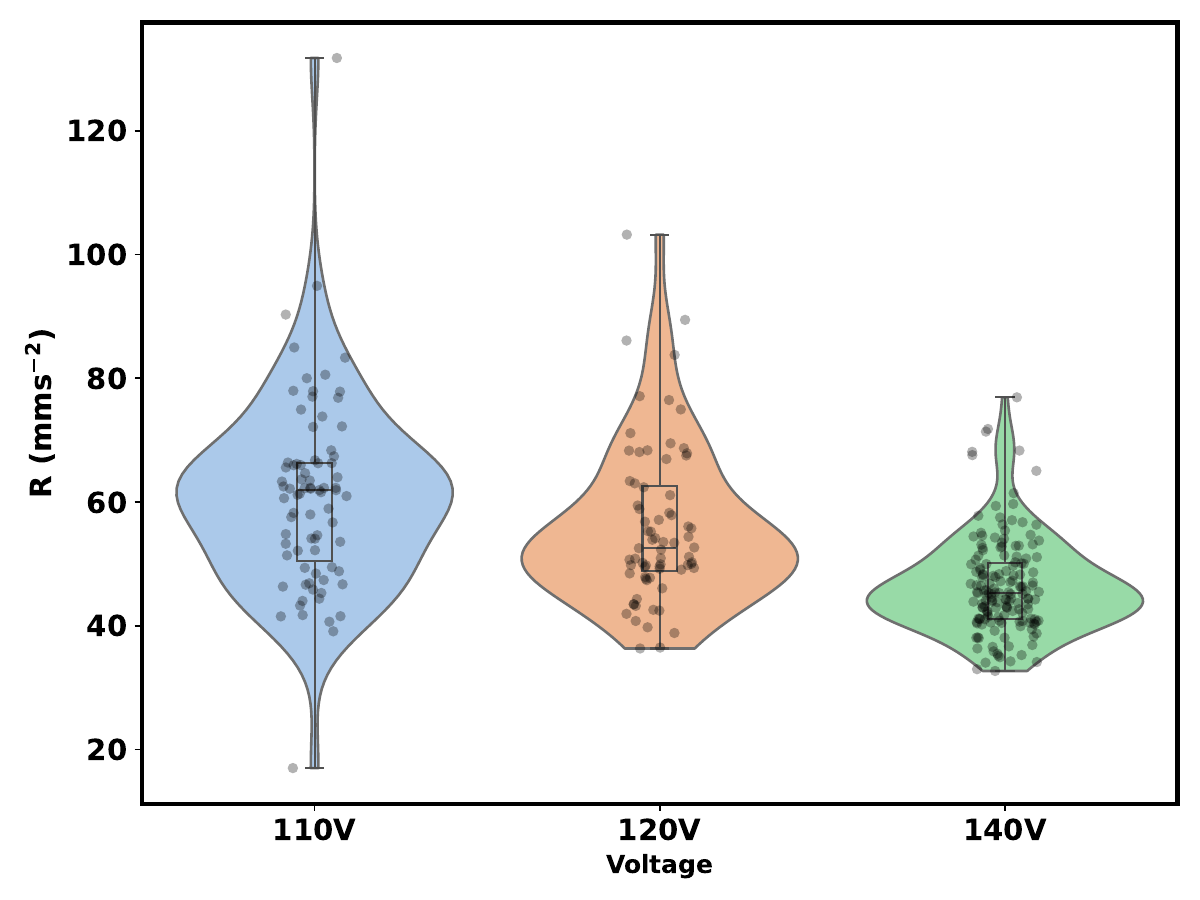}
    \caption{
    Non-reciprocal driving strength $R$ versus bias voltage in the region where melting occurs. Shaded regions show the distribution of $R$ values, while points and error bars denote the spread of the data.
    }
    \label{driving strength}
    \includegraphics[width=1\linewidth]{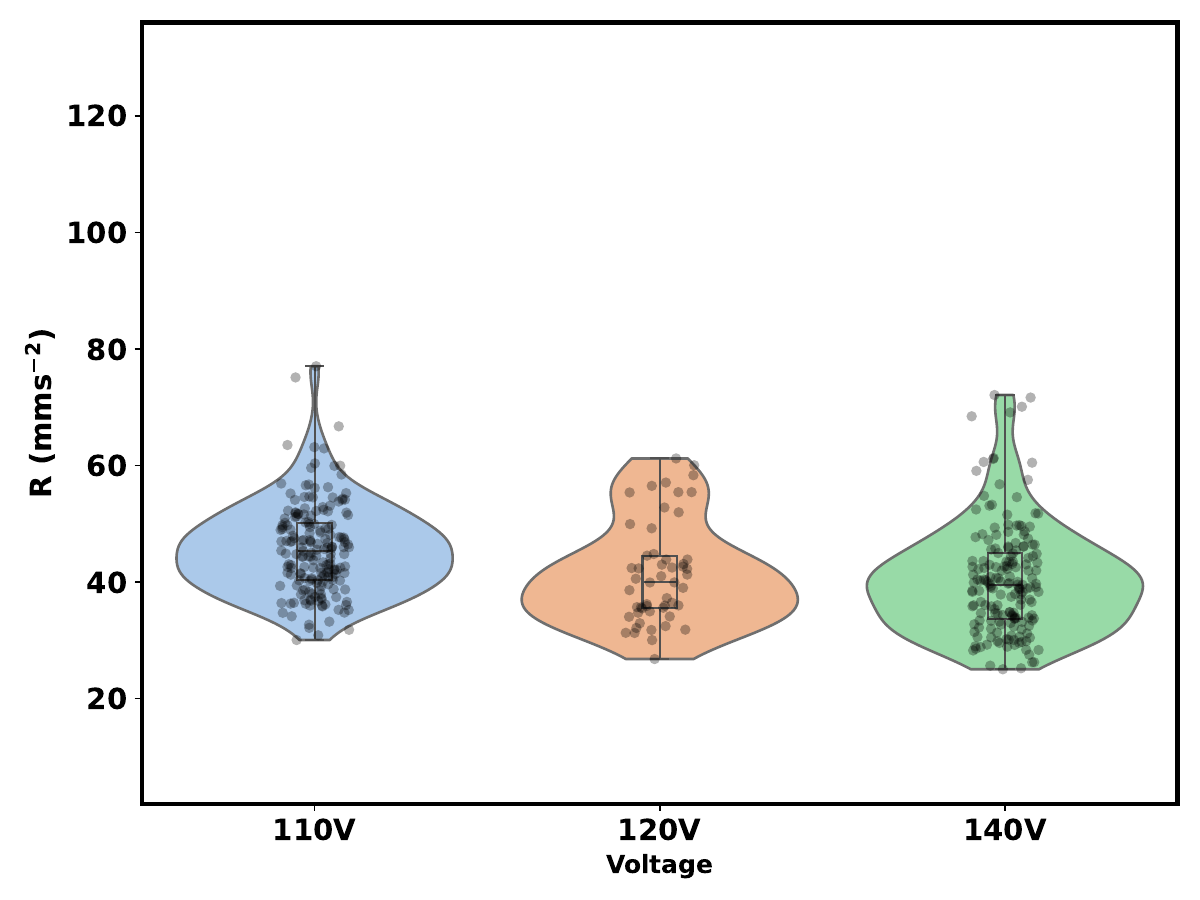}
    \caption{
    Non-reciprocal driving strength $R$ versus bias voltage at the periphery. Shaded regions show the distribution of $R$ values, while points and error bars denote the spread of the data.
    }
    \label{driving strength periphery}
\end{figure}
 We have calculated the value of \textbf{R} for all particle pairs, and the results show a clear dependence on the confinement bias voltage. Fig. \ref{driving strength} and Fig. \ref{driving strength periphery} illustrate how the distribution of \textbf{R} varies with changes in confinement voltage in the central region that is melting and the periphery respectively. In the figures, the shaded regions represent the spread of pairwise non-reciprocal driving strengths at each voltage. A wider shaded region indicates greater variation in \textbf{R} among particle pairs. As the confinement voltage decreases, the spread of the distribution increases for melted region (Fig \ref{driving strength}), indicating enhanced non-reciprocity and variability while the driving strength remains relatively unchanged from the periphery (Fig. \ref{driving strength periphery}).
 
At 140 V, the measured \textbf{R} is small, indicating that interlayer forces are relatively reciprocal and the crystal retains long-range order. Reducing the confinement to 120 V increases interlayer overlap and enhances wakefield-mediated attraction, raising the value of \textbf{R}. This coincides with a marked increase in particle-pairing events. 
At 110 V, where the bilayer collapses into a strongly fluctuating state, the non-reciprocal driving strength reaches its maximum, confirming that asymmetric forces dominate the dynamics at low confinement. 

The monotonic growth of \textbf{R} accompanies the onset of melting and indicates that strengthening non-reciprocal interactions likely contributes to destabilization of the lattice, in combination with confinement-driven modifications of local interaction forces.
\par It is worth mentioning that the neutral gas pressure and discharge voltage were kept constant throughout the experiment; therefore the global plasma parameters remained effectively unchanged. The ring bias modifies only the local sheath electric field in the vicinity of the dust cloud, producing local changes in ion streaming speed, Debye screening and wake-mediated asymmetry. Consequently, the observed changes in \textbf{R} reflect local interaction modifications rather than global plasma reconfiguration.
\par The importance of this observation is underscored by theoretical work by Ivlev et al.\cite{Ivlev2015}, which demonstrates that non-reciprocal interactions fundamentally alter statistical mechanics. Unlike equilibrium Yukawa crystals, where equipartition ensures a well-defined temperature, non-reciprocal systems can sustain different effective kinetic temperatures for different particle species, or even exhibit continuous energy growth with no steady-state equilibrium. Our data support the latter scenario: as non-reciprocity increases, particle pairs act as channels for energy injection from the ion flow into the lattice. The repeated formation and collapse of pairs provides the necessary energy to local rearrangements and velocity fluctuations, ultimately destabilizing the entire crystal.
\par An additional consequence of non-reciprocity is the emergence of a dynamical asymmetry in how instabilities develop. Classical Mode-Coupling Instability (MCI) is triggered when longitudinal and vertical phonon modes intersect, producing resonant energy transfer. 

In contrast, in a bilayer system, the phonon spectra exhibit MCI-like features (frequency shifts, harmonics) but without the expected resonance crossing of longitudinal and vertical modes. This indicates that the instability originates from the combined action of wake-driven non-reciprocity and modified mode coupling, rather than classical MCI alone.
\par In summary, the  analysis of non-reciprocal interactions demonstrates that melting in bilayer dusty plasmas is not only driven by damping thresholds or classical instabilities, but also by a confinement-controlled increase in wake-mediated asymmetry. The monotonic rise of non-reciprocal driving strength with decreasing confinement voltage provides a quantitative, experimentally validated mechanism for how crystalline order gives way to fluid-like coexistence. 

These findings show that increasing non-reciprocal interaction strength is strongly associated with the onset of melting and likely plays a significant role in destabilization, together with other confinement-induced changes in local sheath structure and particle coupling.

\section{\label{sec:level4}Conclusion}
In this work, we explored the mechanisms driving solid-fluid phase coexistence and melting in a bilayer dusty plasma crystal in a DC discharge by combining experimental phonon spectra, particle-level dynamics, and a new quantitative measure of non-reciprocity. Our aim was to investigate the interplay between Mode-Coupling Instability (MCI), particle pairing, and structural disorder under varying confinement conditions.

Phonon spectral analysis revealed MCI-like hotspots near 19 Hz in the acoustic modes, while vertical oscillations in the optical modes occurred in the 30-39 Hz range well above the classical MCI resonance. These results confirm that instability in bilayers differs fundamentally from monolayer MCI, being strongly influenced by interlayer coupling and structural asymmetry. The Schweigert mechanism is also not applicable here: we observed clear vertical oscillations in both layers, growing with decreasing confinement bias, even at constant neutral pressure. This indicates that melting is governed by confinement-driven structural changes rather than a damping threshold.

Side-view imaging revealed repeated formation, collapse, and exchange of particle pairs. These dynamic interactions disrupted local order, redistributed energy, and strongly influenced collective modes. To quantify their impact, we introduced a non-reciprocity metric based on pair accelerations. This analysis showed a clear monotonic increase of non-reciprocal driving strength with decreasing confinement voltage, directly linking microscopic wake-mediated asymmetry to macroscopic instability. Stronger non-reciprocity at low voltages coincided with enhanced particle pairing and eventual melting.

\par Together, these findings point to a hybrid mechanism for melting in bilayer dusty plasma crystals, where vertical confinement, wakefield-mediated non-reciprocal interactions, dynamic particle pairing, and local structural fluctuations act cooperatively to destabilize the crystalline order. These results indicate that dynamic particle pairing amplified by non-reciprocal energy injection is strongly correlated with the phase transition, and accompanies confinement-driven structural and fluctuation effects. The framework developed here, which combines phonon spectral analysis with direct quantification of non-reciprocity, provides a general approach for studying melting, energy transport, and defect dynamics in strongly coupled non-equilibrium systems.

\section{APPENDIX}
%\section*{Supplementary Information}
\begin{figure}[htb!]
    \centering
    \includegraphics[width=\linewidth]{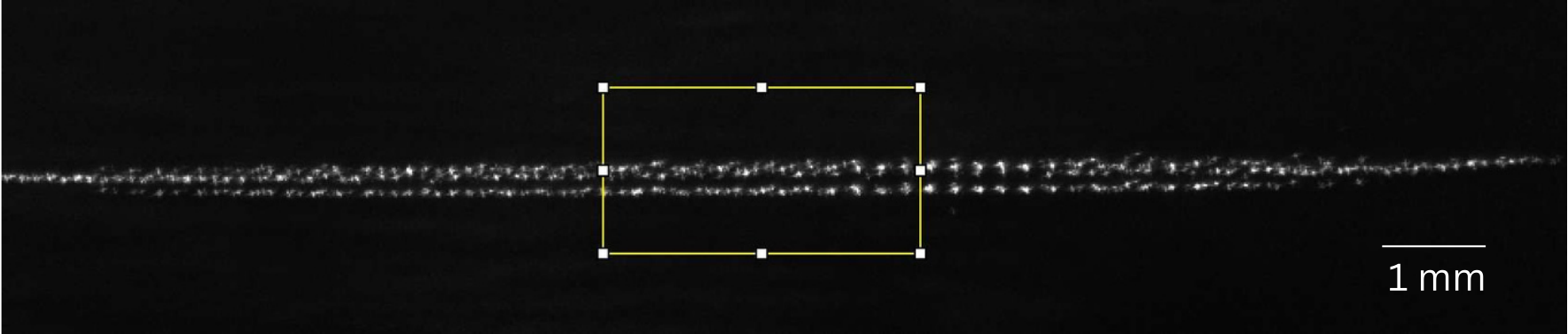}
    \caption{Representative side view data ROI for 140V from which layer profiles were calculated.}
    \label{ROI}
    
\end{figure}

\begin{figure}[htb!]
    \centering
    \includegraphics[width=0.9\linewidth]{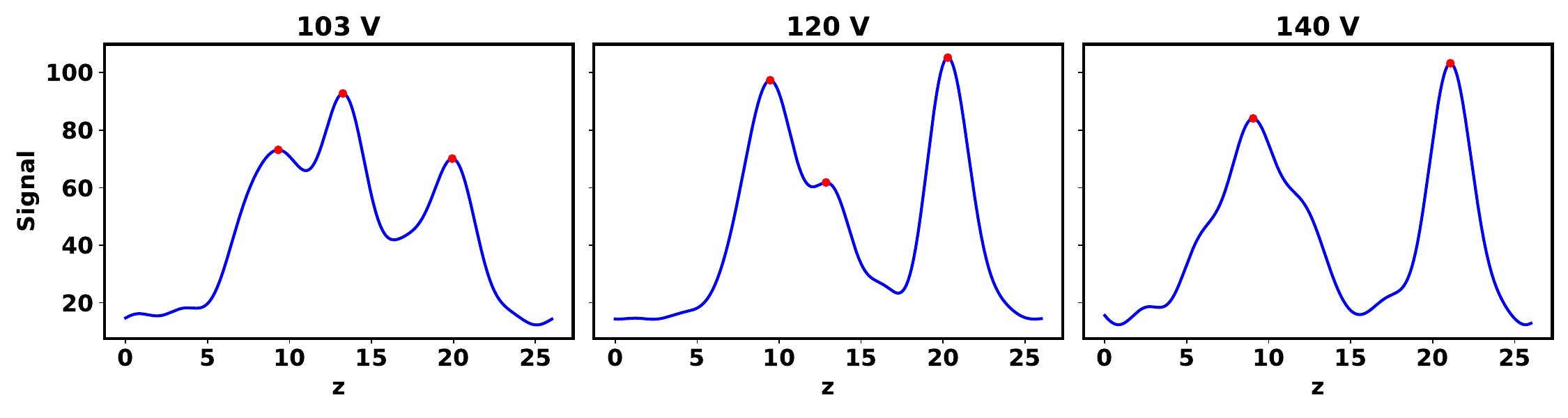}
    \caption{Average vertical intensity profiles corresponding to the ROI in Fig. \ref{ROI}. }
    \label{fig:intensity}
\includegraphics[height=0.5\textheight,keepaspectratio]{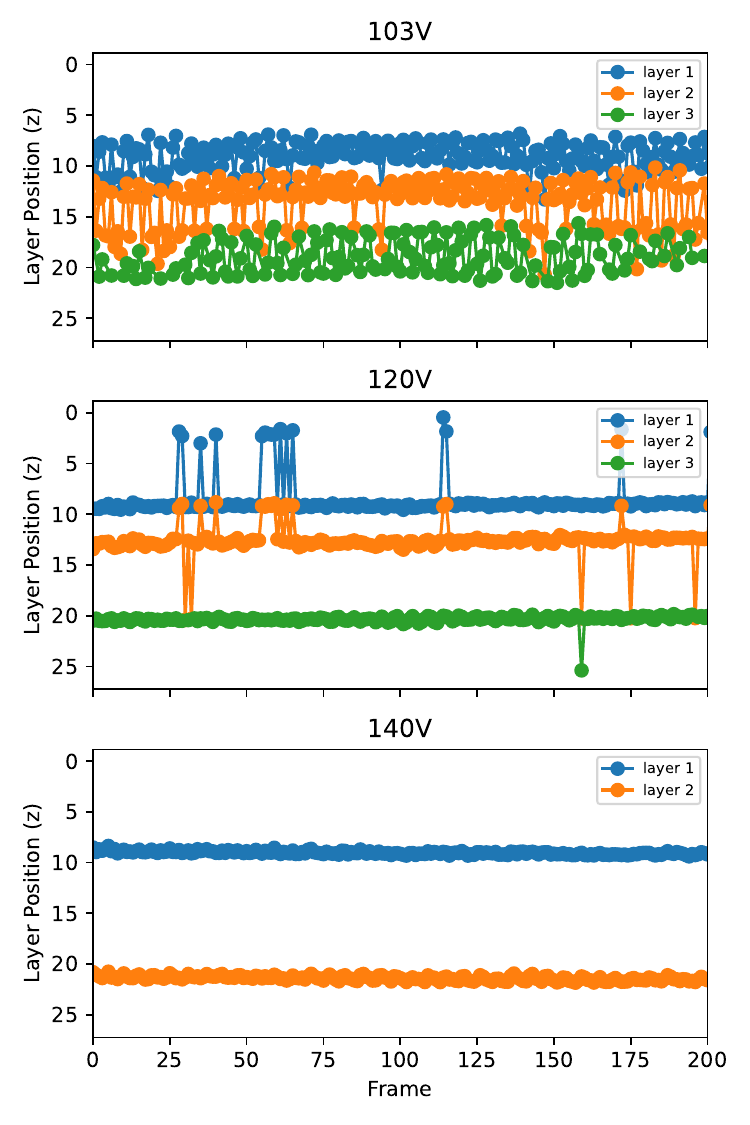}
    \caption{Vertical oscillations of the layers as seen from the side view. We can clearly see the layer interaction increase and the layer move closer as the ring bias voltage is reduced.}
    \label{fig:layer oscillations}
\end{figure}

\subsection*{Analysis Methods}
Each frame of the side-view video (stored as a TIFF stack) was analyzed using \texttt{ImageJ}. For the layer-resolved vertical dynamics shown in Figs. \ref{fig:intensity} and \ref{fig:layer oscillations}, the Region Of Interest (ROI) in Fig. \ref{ROI} was used, and the analysis was performed directly on the image intensity profiles. For each frame, a vertical intensity profile was extracted using the Profile Plot function in ImageJ by averaging the pixel intensity across the horizontal extent of the ROI. The resulting one-dimensional intensity profiles exhibit distinct peaks corresponding to the vertically separated dust layers.
\par The peak positions were identified for each frame, providing the instantaneous vertical locations of the layers. These peak positions were then tracked in time to obtain the vertical trajectories shown in Fig. \ref{fig:layer oscillations}. The profiles shown in Fig. \ref{fig:intensity} correspond to horizontally-averaged vertical intensity profiles for each bias voltage at a fixed frame. Using the same ROI for all figures ensures that the inter-layer separation and layer-resolved oscillations are extracted from identical spatial regions of the dust cloud. 

At a confinement ring bias voltage of 140 V, two well-resolved intensity peaks were consistently observed, indicating a bilayer configuration with a vertical separation large compared to the characteristic screening length, and therefore weak interlayer coupling.

\par At 120 V, the intensity profiles frequently exhibited three distinct peaks, reflecting a splitting of the upper layer into two vertically separated sublayers. The intermediate sublayer appeared within a vertical distance comparable to the effective screening length, resulting in enhanced interaction with both the upper and lower layers, while the direct interaction between the original top and bottom layers remained comparatively weak due to their larger separation.
\par At 103 V, all three intensity peaks shifted closer together, and the separations between adjacent layers decreased substantially. In this regime, the reduced vertical spacing indicates strong interlayer coupling among all layers, consistent with enhanced collective dynamics observed in both side-view and top-view measurements.

\bibliographystyle{ieeetr}
\bibliography{Final_Submission_Bilayer_Complex_Plasma}
\end{document}